\documentclass[runningheads]{llncs}
\usepackage[T1]{fontenc}
\usepackage{amssymb}
\usepackage{footmisc}

\usepackage[ruled, vlined, linesnumbered]{algorithm2e}
\usepackage{listings}
\usepackage[table,dvipsnames]{xcolor}
\lstdefinestyle{mystyle}{
    basicstyle=\ttfamily\scriptsize,
	commentstyle=\color[rgb]{0.35,0.35,0.35}\upshape,
    breaklines=true,
    frame=single,
    tabsize=2,
    rulecolor=\color{black},
    showspaces=false,
    showtabs=false,
    postbreak=\raisebox{0ex}[0ex][0ex]{\ensuremath{\color{gray}\hookrightarrow\space}},
    breakatwhitespace=true,
    numbers=left,
    numberstyle=\small,
    literate={~} {$\sim$}{1},
    upquote=true,
    alsoletter=-,
	morecomment=[l]{\#\ },
    morestring=[b]",stringstyle=\color[rgb]{0,0,0.8},
}

\usepackage{graphicx}
\usepackage{svg}
\usepackage[ruled, vlined, linesnumbered]{algorithm2e}
\usepackage{ulem}
\usepackage{hyperref}
\usepackage{color}

\input{_reviewing}

\begin{document}
\title{RMLStreamer-SISO:
an RDF stream generator from streaming heterogeneous data} 
\titlerunning{RMLStreamer-SISO: an RDF stream generator}
%
\author{Sitt Min Oo\inst{1}\orcidID{0000-0001-9157-7507} \and
Gerald Haesendonck\inst{1}\orcidID{0000-0003-1605-3855}\and
Ben De Meester\inst{1}\orcidID{0000-0003-0248-0987}\and
Anastasia Dimou\inst{2}\orcidID{0000-0003-2138-7972}}
\authorrunning{S. Min Oo et al.}
%
\institute{IDLab, Dept. Electronics \& Information Systems, Ghent University -- imec, Belgium \\
\email{\{x.sittminoo, gerald.haesendonck, ben.demeester\}@ugent.be}
\and
KULeuven, Dept. Computer Science -- Leuven.AI -- Flanders Make, Belgium\\
\email{anastasia.dimou@kuleuven.be}}
\maketitle              
\begin{abstract}

\begin{toremove}
A mature set of stream-reasoning query languages,
such as CQELS and C-SPARQL, exists for querying RDF data streams,
enabling the development of powerful query answering machines
capable of reasoning with semantic data. 
However, nowadays there is still a lack of efficient RDF stream generators to feed these stream-reasoning tools. 
Current state-of-the-art RDF stream generators are non-distributed applications, leading to  inefficient RDF stream generation from high velocity, and large volume of streaming data. 
To efficiently generate RDF stream with scalability in consideration, we extended the RMLStreamer
which was focused on generating RDF from static big data to also 
generate RDF streams from dynamic heterogeneous data streams. 
This paper introduces the RMLStreamer: a
highly scalable solution
to generate RDF streams
with low latency and high throughput from multiple unbounded heterogeneous data sources.
A dynamic window approach is also proposed to handle processing of data from multiple data streams with 
low latency.
An extensive evaluation shows our solution outperforming the related work 
in terms of latency, memory, and throughput. 
The solution is highly scalable both vertically, and horizontally to maintain low latency 
and high throughput processing capabilities. 
\end{toremove}

 Stream-reasoning query languages such as CQELS and C-SPARQL 
 enable 
 query answering over RDF streams.
 Unfortunately, there currently is a lack of efficient RDF 
 stream generators to feed RDF stream reasoners. 
 State-of-the-art RDF stream generators are 
 limited with regard to the velocity and volume of streaming data they can handle. 
 To efficiently generate RDF streams in a scalable way, we extended the RMLStreamer 
to also generate RDF streams from dynamic heterogeneous data streams.
 This paper introduces
 a scalable solution 
 that relies on a dynamic window approach
 to generate RDF streams
 with low latency and high throughput
 from multiple heterogeneous data streams.
 Our evaluation shows that our solution outperforms
 the state-of-the-art by achieving millisecond latency (compared to seconds that state-of-the-art solutions need),
 constant memory usage for all workloads,
 and sustainable throughput of around 70,000 records/s (compared to 10,000 records/s that state-of-the-art solutions take).
 This opens up the access to numerous data streams for integration with 
 the semantic web.
 
\textbf{Resource type:} Software

\textbf{License:} MIT License

\textbf{URL}: \url{https://github.com/RMLio/RMLStreamer/releases/tag/v2.3.0}

\keywords{RML \and Stream processing \and Window Joins \and Knowledge graph generation}
\end{abstract}
\section{Introduction}
An increasing portion of data are continuous in nature, 
e.g., sensor events, user activities on a website, or financial trade events. 
This type of data is known as data streams;
sequences of unbounded tuples
generated continuously
in different rates and volumes~\cite{survey_dsp}. 
Due to the temporal nature of data streams,
low latency computation of analytical 
results is needed to timely react in different use cases, 
e.g., fraud detection~\cite{dsp_edge}. 
Thus,
stream processing engines must efficiently handle
low latency computation
of varying velocity and volume. 

On the one hand,
different frameworks were proposed
to handle data streams,
e.g., Flink, Spark or Storm~\cite{flink,spark,storm}.
On the other hand,
RDF stream processing (RSP) engines,
e.g., CQELS and C-SPARQL~\cite{c-qels,c-sparql,sparql_stream},
were widely studied
and perform high-throughput analysis of RDF streams
with low memory footprints~\cite{c-qels}.
Yet, these stream processing frameworks
are not substantially used in the 
domain of RDF graph generation from streaming data sources,
despite the demand of these mature RSP engines
for more RDF streams.





Between data processing frameworks and stream processing engines,
there are tools to generate RDF streams
from heterogeneous data streams
(e.g. SPARQL-Generate~\cite{sparql-gen},
RDFGen~\cite{rdf-gen}, 
TripleWave~\cite{triplewave},
Cefriel's Chimera~\cite{chimera}). 
However,
some of these tools are inefficient
when the data stream starts to scale in terms of volume and velocity, such as TripleWave, and SPARQL-Generate. While 
other tools are not open sourced nor suitable for the 
mapping of streaming data,
such as RDFGen, and Cefriel's Chimera 
respectively. 
Overall, there are no RDF stream generators
that keep up with the needs of stream reasoning engines
while taking advantage of data processing frameworks
to efficiently produce RDF streams.

In this paper, 
we present the RMLStreamer-SISO,
a parallel, vertically and horizontally scalable stream processing engine 
to generate RDF streams from heterogeneous data streams of 
any format (e.g. JSON, CSV, XML, etc.).
We extended previous preliminary work~\cite{rmlstreamer-big-data}
of heterogeneous data
stream mapping solution:
an open source implementation on top of Apache Flink~\cite{flink}, 
available under MIT license, 
which generates high volume RDF data
from high volume heterogeneous data.
RMLStreamer-SISO extends RMLStreamer
to also support any input data streams and export RDF streams
(Stream-In-Stream-Out (SISO)).
RMLStreamer-SISO now supports a much larger part of the RML 
specification\footnote{Implementation report of RML:~\url{https://rml.io/implementation-report/}},
including all features of RML but relational databases.

The RMLStreamer-SISO outperforms the the state-of-the-art tools 
when handling high velocity data stream,
increasing the throughput it could handle
while maintaining low latency.
The RMLStreamer-SISO achieves millisecond latency,
as opposed to seconds that state-of-the-art solutions need,
constant memory usage for all workloads,
and sustainable throughput of around 70,000 records/s,
compared to 10,000 records/s that state-of-the-art solutions take.



Through the utilization of a low-latency tool like RMLStreamer-SISO, 
legacy streaming systems could exploit the unique characteristics of real-life streaming data,
while enabling analysts to exploit the semantic reasoning using knowledge graphs in real-time and 
have access to more reliable data.

The contributions presented in this paper are:
(i)~an algorithm to generate the RDF streams
from heterogeneous streaming data;
(ii)~its implementation, the RMLStreamer-SISO,
as an extension of RMLStreamer; and
(iii)~an evaluation demonstrating that the RMLStreamer-SISO 
outperform the state-of-the-art.
The paper is structured as follows:
Section~\ref{sec:relWork} discusses related work,
Section~\ref{sec:Methodology} the approach and its implementation,
Section~\ref{sec:evaluation} the evaluation of RMLStreamer-SISO
against state-of-the-art,
Section~\ref{sec:results} the results of our evaluations, and 
Section~\ref{sec:conclusions} concludes our work with possible future works. 

\section{Related Works}
\label{sec:relWork}


Streaming RDF mapping engines transform heterogeneous data streams
to RDF data streams. Several solutions exist in the literature 
for generating RDF from persistent
data sources~\cite{Simsek2019RocketRMLA,sdm-rdfizer,rmlstreamer-big-data,chimera_ontop}, 
but only few generate RDF from data streams \cite{rdf-gen,sparql-gen,triplewave}. 
Although the implementations details are elaborated in these works, 
their evaluations are designed 
without considering the different data stream behaviours 
nor the resource contention between different
evaluation components.

\textbf{TripleWave}~\cite{triplewave} generates RDF streams from streaming or static data sources
using R2RML mappings, and publishes them as RDF
stream. However, the R2RML mappings of 
TripleWave are invalid according to the 
specifications of R2RML and it does not support joins.
Although it is purported to support several input sources, the user 
has to write the code to process the input data and iterate 
over them before using the tool.
This can result in poor performance from 
improper implementation. 
Last, it is not designed to support
distributed parallel processing, resulting in limited scaling with
data volume and velocity.

\textbf{RDF-Gen~\cite{rdf-gen}} generates static or streaming RDF data
from static or streaming data sources. A Data connector
communicates with the data source, iterates over its data
entries, and converts every entry to a record of values. These
records are converted to RDF using a graph template: a
listing of RDF-like statements with variables
bound to the record values coming from data connectors. RDF-Gen
generates RDF on a per record basis, theoretically allowing
a distributed parallel processing set-up. However, the current
implementation and documentation show no indication of a
clustered setup nor how to run it.

\textbf{SPARQL-Generate}~\cite{sparql-gen} extends SPARQL 1.1 syntax to support mapping 
of heterogeneous data to RDF data.
SPARQL-Generate could be 
implemented on top of any SPARQL query engine, and knowledge engineers with SPARQL experience could use it with ease. The reference implementation of 
SPARQL-Generate\footnote{SPARQL-Generate:~\url{https://github.com/sparql-generate/sparql-generate}}
generates RDF streams from data streams, even though it is not reported 
in the original paper.
Although joining data from multiple sources is supported, 
SPARQL-Generate waits for one of the data streams 
to end first before 
consuming other data sources to join the data. Thus, joins with unbounded
streaming data sources are not supported. 
The implementation is based on single machine setup without scaling with data volume and velocity.

\textbf{Cefriel’s Chimera}~\cite{chimera} is  an integration framework based on 
Apache Camel~\footnote{Apache Camel: \url{https://camel.apache.org/}}
split into four “blocks” of components to  map heterogeneous data to RDF data: lifting block, data enricher, inference enricher, and lowering block. 
Chimera aims to be modular and allows each block to be
replaced with custom implementations.
The current implementation
uses a modified version of RMLMapper\footnote{RMLMapper: \url{https://github.com/RMLio/rmlmapper-java}} in the lifting 
block for data stream processing. However, the whole RML mapping process 
is recreated with each incoming message which could lead to high performance overhead 
in a highly dynamic data stream environment.

\section{Stream In - Stream Out (SISO)}
\label{sec:Methodology}

We extend RMLStreamer's architecture
for generating RDF from persistent big data
sources~\cite{rmlstreamer-big-data} to also generate RDF streams from
heterogeneous data streams with high data velocity and volume,
while keeping the latency low.
The RDF mapping language (RML)~\cite{rml}, a superset of R2RML, 
expresses customized mapping from heterogeneous data sources to 
RDF datasets.
We illustrate the concepts of RML with the example RML document in Listing~\ref{lst:rml}. 

\begin{figure}[!htbp]
    \centering
    \includegraphics[width=\linewidth]{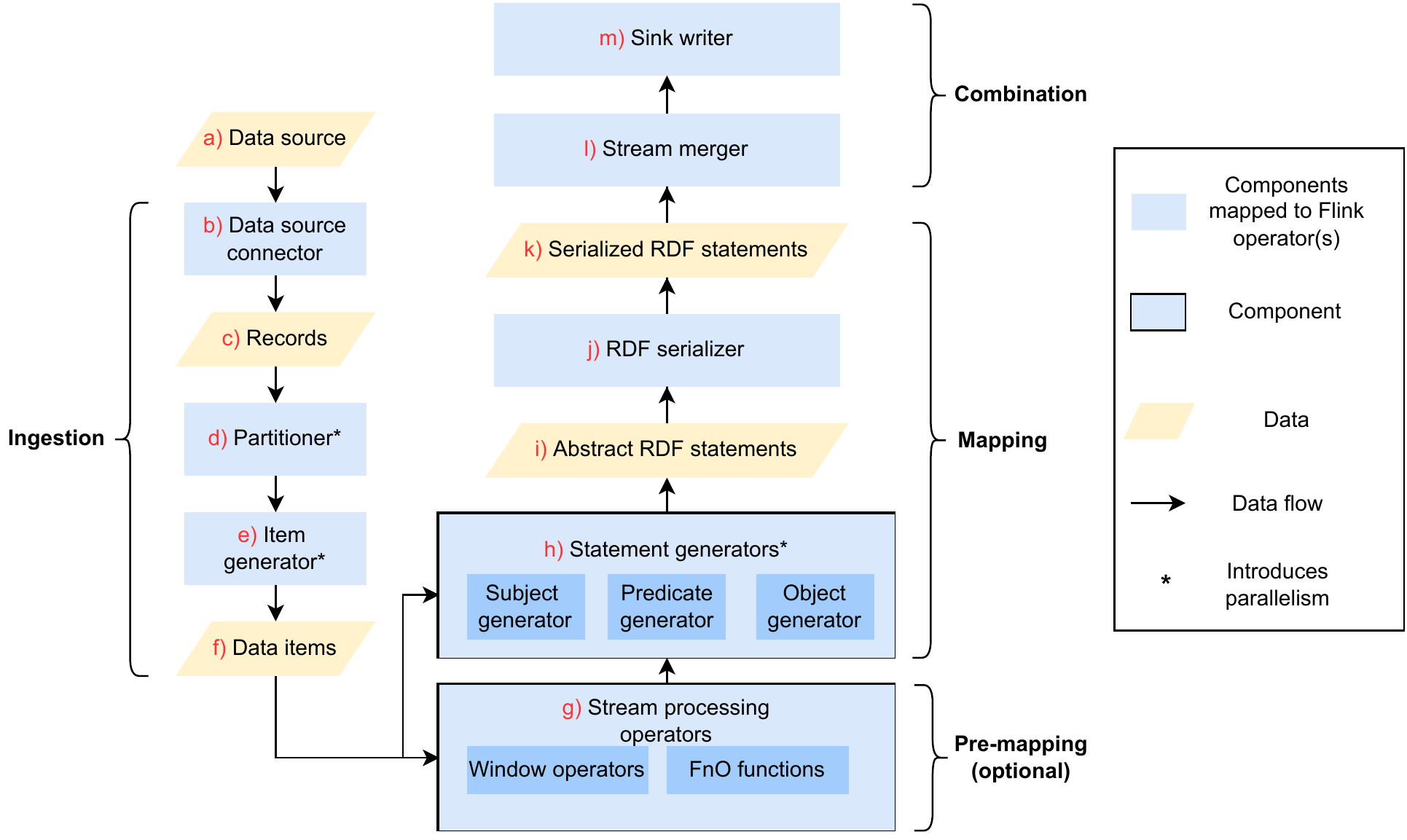}
    \caption{Workflow of RMLStreamer. Data flows from the \textit{Data Source} at the top 
    through all the components pipeline to the \textit{Sink writer} at the bottom.}
    \label{fig:rml_architecture}
\end{figure}

We break the process of generating RDF from a data stream into tasks and subtasks (Figure \ref{fig:rml_architecture}).
Each task or subtask is a stream processing operator acting on an incoming data stream. 
They could be chained one after the other to form a pipeline of operators and result 
in one or more outgoing data streams. 
This approach introduces parallelism on both data and processing level, 
enabling each data stream and operator 
to be processed and executed respectively in parallel.

To illustrate 
RMLStreamer-SISO’s pipeline, 
we use the examples in Listing \ref{lst:json} and \ref{lst:rml}. The
mapping document in Listing~\ref{lst:rml} is used to join and 
map JSON data (Listing~\ref{lst:json}) from
websocket streams to RDF with dynamic window join. 

\begin{lstlisting}[
label=lst:json,
caption=Data records from 2 data streams 
``Flow'' \& ``Speed''.]
// data records from Speed stream
{"speed":123.0,"time":"14:42:00","id":"lane1"}
// data records from Flow stream
{"flow":1680,"time":"14:42:00","id":"lane1"}
\end{lstlisting}

\begin{lstlisting}[label=lst:rml,caption=Example RML Mapping file to generate streaming RDF from the streaming heterogeneous data of Listing \ref{lst:json}.]
# prefix definitions omitted
_:ws_source_ndwSpeed a td:Thing ;
  td:hasPropertyAffordance [ td:hasForm [
    hctl:hasTarget "ws://data-streamer:9001" ; # URL and content type
    hctl:forContentType   "application/json" ; # Data format
    hctl:hasOperationType "readproperty" ] ] . # Read only
_:ws_source_ndwFlow a td:Thing;
  td:hasPropertyAffordance [ td:hasForm [
    hctl:hasTarget "ws://data-streamer:9000" ;
    hctl:forContentType   "application/json" ;
    hctl:hasOperationType "readproperty" ] ] .
<JoinConfigMap> a rmls:JoinConfigMap ;
  rmls:joinType rmls:TumblingJoin .          # Trigger/eviction type
<NDWSpeedMap> a rr:TriplesMap ;
  rml:logicalSource [                        # Describes data source
    rml:source _:ws_source_ndwSpeed ;
    rml:referenceFormulation ql:JSONPath ;   # JSONPath iterator 
    rml:iterator "$" ] ; # Iterates the data as JSON root object
  rr:subjectMap [        # Generation of the subject IRI
    rr:template "speed={speed}&time={time}" ] ;
  rr:predicateObjectMap [ # Describes how predicate and object are generated
    rr:predicate <http://example.com/laneFlow> ;
    rr:objectMap [ 
      rr:parentTriplesMap <NDWFlowMap> ;     # TripleMap to be joined with 
      rmls:joinConfig  <JoinConfigMap> ;     # Configuration of join window
      rmls:windowType  rmls:TumblingWindow ; # Type of join window
      rr:joinCondition [ # Attributes on which the data records are joined
        rr:child "id" ; rr:parent "id" ; ] ] ] .
<NDWFlowMap> a rr:TriplesMap ;
  rml:logicalSource [
    rml:source _:ws_source_ndwFlow ;
    rml:referenceFormulation ql:JSONPath ;
    rml:iterator "$" ] ;
  rr:subjectMap [ rr:template "flow={flow}&time={time}" ] .
\end{lstlisting}

\subsection{RDF stream generation workflow}
\label{ssec:window}

Our approach consists of a workflow with four tasks (see Figure \ref{fig:rml_architecture}):

\paragraph{Ingestion} The ingestion task captures data streams and prepares the data records for the mapping task. Each data stream triggers one ingestion task that can run in parallel with the other ingestion tasks spawned by the other data streams. The ingestion task can be divided in three subtasks:
    \begin{enumerate}
        \item \textit{Data source connector (Figure \ref{fig:rml_architecture}, \textcolor{red}{(b)})}: This subtask is responsible for
        connecting to a (streaming) data source \textcolor{red}{(a)}. It reads data records from the source and passes 
        these records \textcolor{red}{(c)} on to the stream partitioner.
        
        \item \textit{Stream partitioner \textcolor{red}{(d)}}: The stream of data records is optionally partitioned 
        in disjoint partitions to be fed to the next subtask. The partitioning depends 
        on the order's maintenance.
        If the exact order of the incoming 
        data records is not important to be maintained, then these records can be
        distributed evenly among multiple instances of the next subtask, increasing 
        parallelism. If the order of generating RDF statements needs to correspond 
        with the order of the incoming data records, then the stream is not distributed at this stage. 
        
        \item \textit{Item generator \textcolor{red}{(e)}}: One data record can lead to zero or more RDF statements. 
        This subtask splits a data record in zero or more items of internal
        representation called \textit{data items} \textcolor{red}{(f)},
        according to the logical iterators defined in the mapping document, before the actual 
        mapping task takes place. Using the sample data and the mapping document from 
        Listing \ref{lst:json} and \ref{lst:rml} respectively, this subtask will use the logical iterator `\texttt{\$}', 
        a JSONPath\footnote{JSONPath documentation: \url{https://goessner.net/articles/JsonPath/index.html}}, 
        to generate data items from each data record shown in Listing \ref{lst:json}. 
        In this case, the logical iterator is the JSON root object, so the data item 
        is the same as the incoming data records. Otherwise, if the data
        record contains a list of sub-records, and the logical iterator is 
        specified over the list (e.g., \texttt{\$.list[*]}), each of these sub-records
        are turned into \textit{data items}.
    \end{enumerate}
\paragraph{Pre-mapping (optional)}
    Before the data items are mapped to RDF, 
    the data items may be processed with custom data transformations 
    defined with FnO~\cite{fno},
    or the window operators,
    such as joins, aggregates, and reduce.
    The FnO functions could be as simple as changing
    letters to uppercase or as complex as the window joins. 
    This stage is optional and omitted if the RML document does not define pre-mapping functions. 
    The pre-mapping task \textcolor{red}{(g)} is right before the mapping task
    since the data fields requiring preprocessing 
    can be more than the data fields needed for mapping to RDF data.
    For example, with the given inputs and mapping document (Listing \ref{lst:rml}), the data items (Listing \ref{lst:json})
    from the two input streams, ``\texttt{Flow}'' and ``\texttt{Speed}'', are first buffered inside a window,
    and then joined based on their \texttt{internalId} value.
    Data records, having the same value for the ``\texttt{id}'', are joined pairwise.
    If windows joins were implemented after the mapping stage,
    the verbosity of RDF would substantially increase 
    the network bandwidth.
    More, to fully map the data before joining,
    RMLStreamer-SISO needs to know all 
    attributes present in the raw data records which would be 
    infeasible. 
    
    To support joining with windows, RML was extended.
    New vocabulary terms were defined to support windowing operations with RML.
    We defined two new properties:
    \textit{rmls:windowType} 
    to provide the type of window to be used when joining
    and \textit{rmls:joinConfig} 
    when joining the \textit{Child} and \textit{Parent Triple Map}
    to define how the trigger, and eviction are fired inside the window. 
    
    Section~\ref{ssec:window} details the dynamic windowing algorithm and Section \ref{sec:Implementation} elaborates on the design choice and windows' implementation for RMLStreamer-SISO.
    
\paragraph{Mapping} RDF statements are generated from data items coming from the ingestion task and the pre-mapping task.
    \begin{enumerate}
        \item \textit{Statement generator \textcolor{red}{(h)}}: Each data item leads to one or more RDF statements
        in this sub task. Each statement is generated in parallel as an abstract RDF 
        statement \textcolor{red}{(i)} which could be fed to the next subtask for serialization. 
        
        \item \textit{RDF serializer \textcolor{red}{(j)}}: 
        The abstract RDF statements are serialized into various 
        RDF serialisations based on the configuration given to the RMLStreamer. 
        
    \end{enumerate}
    
\paragraph{Combination} This task brings back together all streams of RDF statements \textcolor{red}{(l)} into one final RDF stream which will be written
    out using the sink writer \textcolor{red}{(m)}.

\subsection{Heterogeneous data streams join in RDF streams}
\label{ssec:window}

Supporting \textit{joins} in RMLStreamer-SISO
and any streaming RDF generator, is not trivial
as windowing techniques are required
for unbounded and unsynchronized streaming data. 
Unlike batch processing where data is bounded, 
processing whole data streams in memory is unsustainable 
due to the continuous and infinite characteristics of streaming data.
Therefore, stream processing engines use buffers called \textit{windows} to hold the most
recent stream of records in memory.
The windows' lifetime is measured in terms of time interval, thus,
the \textit{window interval} determines the size of the window
and their operation behaviour
is defined by the trigger, and the eviction events~\cite{generic}. 
A \textit{trigger 
event} occurs when an operator is executed to process the data 
records inside the window interval. 
An \textit{eviction event} occurs 
when the window evicts the data records inside its buffer. 
\begin{figure}
    \centering
    \includegraphics[width=\linewidth]{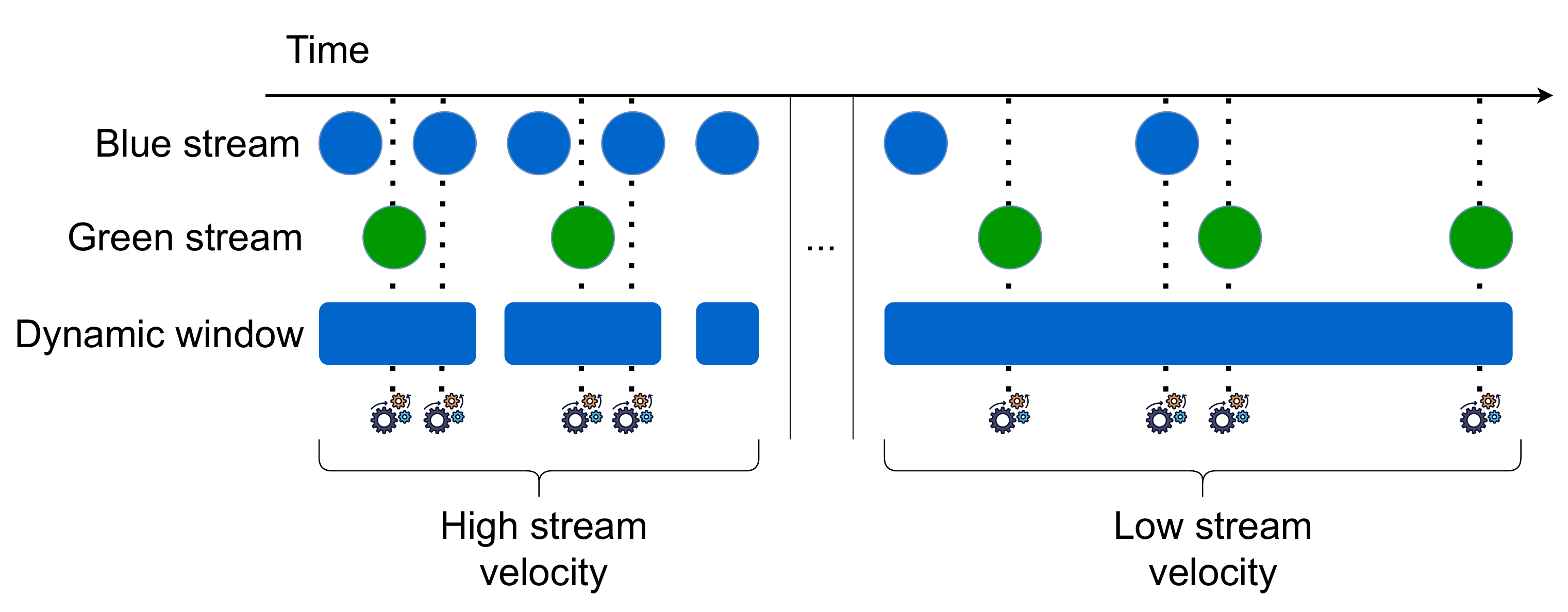}
    \caption{Behaviour of the dynamic window under high, and low stream velocities.
    The cogwheels are the \textit{trigger} events representing the
    moment when the data records are processed. In this figure, 
    the \textit{trigger} events are fired with every new data record,
    and 
    only when there is at least one data record from each data stream. 
}
    \label{fig:dynamic}
\end{figure}

We opted for an eager 
trigger implementation to lower the latency of RMLStreamer’s responses for the windowed joins' implementation. 
The joined results are emitted as soon as possible without waiting for the eviction event to occur.
We designed a dynamic window which adapts its window 
intervals according to the velocity of the incoming data streams.
Adaptive windowing~\cite{adaptive} was studied in the context of batch 
stream processing with a positive impact on the stream processing job's 
performance: lower latency, and higher throughput. 
We opted for a simple cost metric based on the data 
records' number to keep the memory and latency low in a real-time
stream processing environment where the time constraint is
more stringent.

The algorithm is inspired by the additive-increase, 
and multiplicative 
decrease algorithm of TCP congestion control~\cite{chiu1989analysis}.
Figure \ref{fig:dynamic} shows 
the high level behaviour of our dynamic window for the two different stream velocities. 
When the data stream velocity is high, the window size shrinks to process the data 
records as fast as possible, keeping the latency low and throughput high.
When the data stream velocity is low,
the size of the window grows to wait for more data records and process them.
This ensures that the window do not miss the records due to short window 
size. 
We elaborate the details of the algorithm below.
For each window, the following configuration parameters are provided: 
\begin{enumerate}
    \item $|W|$: The window interval
    \item $\epsilon_u$  \&  $\epsilon_l$: Upper and lower threshold limit for total cost metric
    \item $U$ \& $L$: Upper and lower limit for the window interval
    \item $Limit(List_P)$ \& $Limit(List_C)$: Upper limit size for parent and child stream
\end{enumerate}

\begin{algorithm}[htbp]
    \DontPrintSemicolon
    \KwData{$|W|, \epsilon_u, \epsilon_l, U, L, Limit(List_P),Limit(List_C), S_P, S_C$}
    $cost(List_P) = |S_P| / Limit(List_P)$  
    
    $cost(List_C) = |S_C| / Limit(List_C)$     

    total cost $ m = cost(List_P) + cost(List_C)$  
  
  \tcp{adapts window size based on cost}
    \If{$m > \epsilon_u$} 
    {
        $|W| = |W| / 2 $ 
        
        $Limit(List_P)  = Limit(List_P) * cost(List_P) * 1.5$  

        $Limit(List_C)  = Limit(List_C) * cost(List_C) * 1.5$  
    }
    \ElseIf{$m < \epsilon_l$}
    {
        $|W| = |W| * 1.1$ 

        $Limit(List_P)  = Limit(List_P) * cost(List_P) * 1.5$  

        $Limit(List_C)  = Limit(List_C) * cost(List_C) * 1.5$  
    }

    clean both $List_C$ and $List_P$
    
    clip $|W|$ in the range of $[L,U]$
    \caption{Dynamic window $onEviction$ routine}
    \label{alg:dynamic_eviction}
\end{algorithm}

Since we implement the join operator with eager execution,
the trigger event is fired when the current record $r_c$ arrives 
inside the window. We denote the current window as $W$ with interval size 
$|W|$. The streams are denoted as $S_p$ and $S_c$ with the 
corresponding states $List_p$ and $List_c$, for the parent and child stream respectively 
(the parent and child stream follows the RML specification for joining triples maps).
The states contain the records
from their respective streams inside the window with for example $|S_p|$ 
denoting the number of records from $S_p$.
$List_p$ and $List_c$ are only used in cost calculation to determine 
if the window interval needs to be changed; they do not limit 
the amount of records that could be buffered 
inside the window. 

At each eviction trigger, we calculate 
the cost for each list states $List_p$ and $List_c$. For example, the cost for 
$cost(List_P) = |S_p|/Limit(List_P)$.
The total cost is $m = cost(List_p) + cost(List_c)$ and it is
checked against the thresholds $\epsilon_l$ and $\epsilon_u$ to adjust the 
window interval accordingly. We assume the stable zone to
be achieved if the total cost fulfils the predicate: 
$\epsilon_l \leq m \leq \epsilon_u$. Algorithm~\ref{alg:dynamic_eviction} shows the pseudo-code 
for the eviction algorithm we just elaborated.

\subsection{Implementation}
\label{sec:Implementation}
RMLStreamer-SISO is released as version 2.3 of RMLStreamer 
to utilize Flink's parallelism for 
horizontal scaling (via distributed processing in a network and 
vertical scaling (via multi-threaded execution of tasks).
The update brings the windowing support 
for joining multiple data streams, the dynamic windowing algorithm,
and FnO~\cite{fno} as an extension point for joins execution. 
The code and usage instructions 
for RMLStreamer-SISO are available 
online at the Github repository: \url{https://github.com/RMLio/RMLStreamer}.

Windowing support is implemented through the use of Flink's windowing 
API\footnote{Window: \url{https://nightlies.apache.org/flink/flink-docs-release-1.14/docs/dev/datastream/operators/windows/}}
for common types of window, e.g., Tumbling Window. 
We implemented the \textit{KeyedCoProcessFunction} provided by Flink's low-level stream processing API
to manage the different states required 
for the algorithm (Algorithm~\ref{alg:dynamic_eviction}) of the dynamic window.
We implemented the dynamic windowed join before the mapping stage, to group input streams and reduce network bandwidth usage. 
The generated RDF stream
could be windowed by the RDF stream processing engines consuming the output.

Currently, FnO functions jar files have to be
compiled together as part of the RMLStreamer-SISO jar. 
Examples on 
github\footnote{RMLStreamer-SISO: 
\url{https://github.com/RMLio/RMLStreamer}} show 
the working of RMLStreamer-SISO with TCP 
data stream. We also provide an extensive documentation 
on RMLStreamer-SISO in a containerized environment with docker\footnote{Docker: \url{https://docker.com}}.

\section{Evaluation}
\label{sec:evaluation}
An extensive evaluation was conducted
focused on variable data stream velocity, volume and variety of data formats 
to emulate the real-life workloads as close as possible.
The code for the evaluation is available on github\footnote{Benchmark: \url{https://github.com/s-minoo/rmlstreamer-benchmark-rust}}.
Since RMLStreamer-SISO is situated between traditional stream processing and RSP,
state-of-the-art 
approaches for benchmarking in these domains are combined: architectural design 
of RSPLab~\cite{rsplab}, workload design of Open Stream 
Processing Benchmark \cite{dsp_eval}, and 
measurement strategies of Karimov et. al \cite{karimov_2018}. 

We compare the RMLStreamer-SISO with the state-of-the-art streaming 
RDF generator, SPARQL-Generate, 
which is actively maintained, used and supports the same features 
as RMLStreamer-SISO.
The other tools were not considered for different reasons:
TripleWave requires a custom implementation to process each data stream
and feed it in TripleWave which means it cannot be used as-it-is.
More, TripleWave is meant purely for feeding RDF streams to RDF stream processing engines
without performing joins, 
therefore it would have been an unfair comparison both in terms of features and scope.
RDF-Gen's source code is not available, but only a jar is available without any instructions to run it.
Both TripleWave and RDF-Gen are also not actively maintained. 
Finally, Cefriel's Chimera restarts the RDF mapping engine with every data record,
which means that the processing of the input and mapping is not performed in a true streaming manner;
the comparison would not be meaningful.

\paragraph{Data source}
\label{ssec:data_source}
The input data used in the evaluation comes from time annotated 
traffic sensor data from the Netherlands provided by 
NDW (Nationale Databank Wegverkeersgegevens)\footnote{NDW: \url{https://www.ndw.nu/}}, 
and also used by Van Dongen et al~\cite{dsp_eval}.
It contains around 68,000 rows of
CSV data with two different measurements across different lanes on a highway:
number of cars (flow), and their average speed.
The two measurements are streamed 
through a websocket data streaming server. 

\paragraph{Metrics}
\label{ssec:metric}

Stream processing frameworks are typically evaluated using
two main metrics: latency and throughput~\cite{karimov_2018}. 
Latency can be further distinguished into two types: processing-time 
latency, and event-time latency. 
\textit{Processing time latency} 
is the interval between the data record’s arrival time at the
input and the emission time at the output of the streaming 
engine~\cite{karimov_2018}.
\textit{Event-time latency} is the interval between the creation
time, and the emission time at the streaming engine’s 
output, of the data record~\cite{karimov_2018}.  
Latency measurement requires to consider the effect of coordinated occlusion, where
the queueing time, a part of the event-time, is ignored~\cite{karimov_2018}.
Therefore, we consider event-time latency as our latency 
measurement to take the effect of coordinated occlusion in
consideration. 

For our evaluation, we considered the event-time latency of 
each record,
the throughput as number of consumed records per second,
the memory and CPU usage of the engine’s docker container.
The measurements are captured on a machine separate from the 
host machine of the System Under Test (SUT), where memory and CPU usage are measured using cAdvisor%
\footnote{\label{cadvisorft}cAdvisor: \url{https://github.com/google/cadvisor}}.
By treating the SUTs as a blackbox,
we ensure that the measurement of the metrics incurs
no performance penalty nor resource contention with 
the SUTs during the evaluation.

\begin{figure*}[!htbp]
    \centering
    \includegraphics[width=\textwidth]{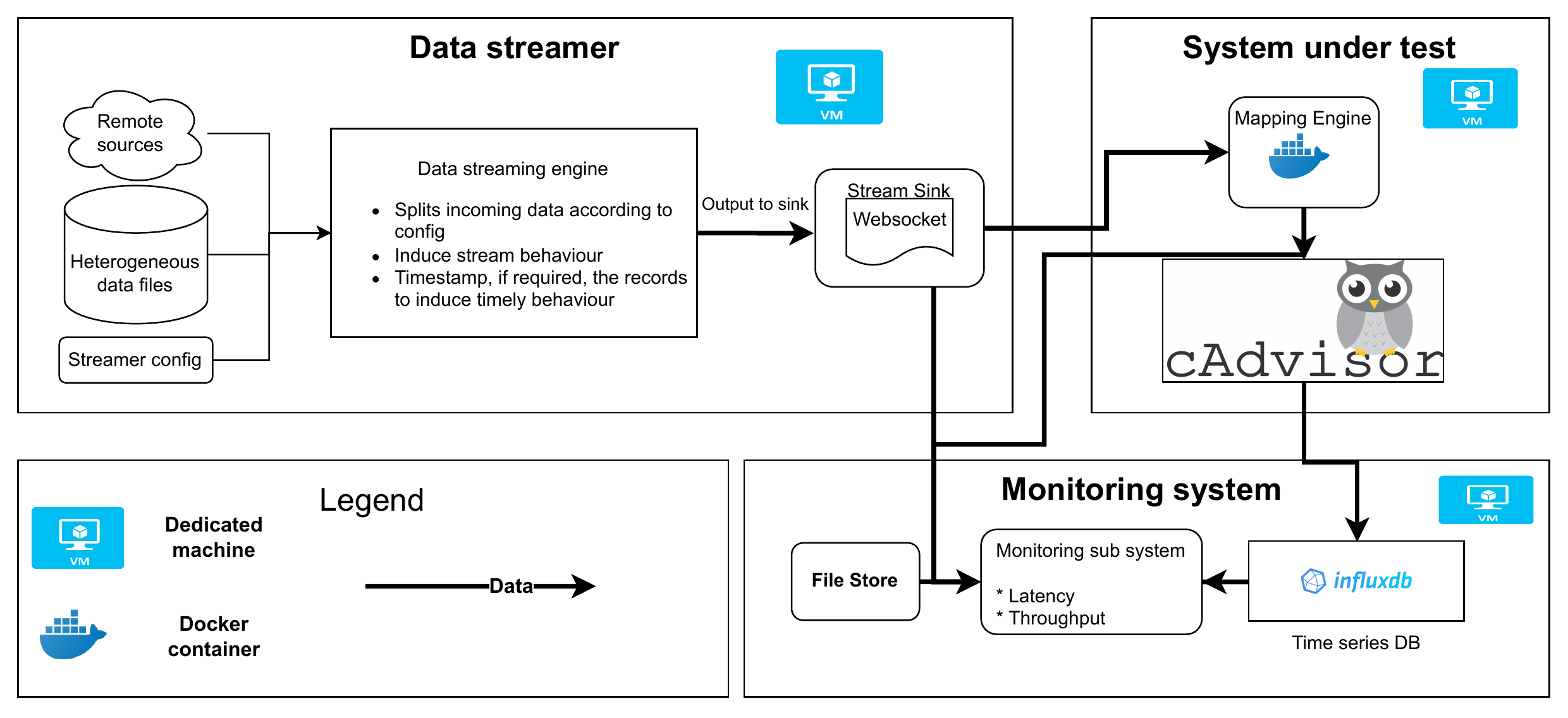}
    \caption{Benchmark architecture to evaluate the different engines, inspired by RSPLab. }
    \label{fig:eval_architecture}
\end{figure*}

\paragraph{Evaluation set up}
The architectural design is a modification of RSPLab 
with a custom data streaming component (Figure \ref{fig:eval_architecture}).
It consists of three components: 
a) the data streamer, 
b) the system under test, and 
c) the monitoring system.

With the proposed architecture where each components is isolated, 
we aim to reduce the influence of the benchmark components on the engine 
during the evaluation process.
The modularity of the setup also
increases the flexibility of configuring the evaluation environment 
with minimal changes for the engines.

\paragraph{Workload Design}
\label{ssec:workload}
To evaluate the performance of the engines under different data 
characteristics and processing scenarios,
we devise three different workloads:
(i) throughput measurement, (ii) periodic 
burst, and (iii) scalability measurement. 
As SPARQL-Generate is unable to
join unbounded streaming data (it expects data streams with an end, Section~\ref{sec:relWork}),
we evaluated the two workloads (throughput measurement and periodic burst)
without joining functionality to compare. 

\begin{itemize}
    \item throughput measurement: the data stream throughput is constant and steadily increases with each run
    to determine the engine's \textit{sustainable throughput}~\cite{karimov_2018}. 
    CPU, latency and memory usage are measured. 
    \item periodic burst: a burst of data records is emitted periodically to mimic fluctuations in data streams; CPU, memory, latency and throughput are measured. 
    \item scalability measurement: RMLStreamer-SISO is evaluated in two modes: centralised mode without parallelism and distributed mode with parallelizable data to measure the impact of parallelism on its scalability. In both modes, data from two input streams are joined and latency is measured. 
\end{itemize}

\paragraph{System specifications}
We ran the evaluation on a single machine with multiple docker containers 
to emulate the communication between the data streaming source and the mapping 
engine in a streaming network environment. The machine has Intel i7 CPU with
8 cores at 4.8Ghz, 16GB RAM, and 200GB hard disk space.
The data streamer and the monitoring system docker containers
(Figure~\ref{fig:eval_architecture}) have access to 4 of the cores, and the 
SUT docker container has access to the leftover 4 cores. This prevents CPU resource
contention between the SUT and the other components used for running the
evaluation. 

To evaluate horizontal scaling, 
the data streamer component 
is replaced with Apache Kafka to support parallel ingestion of data streams by 
RMLStreamer-SISO. Apache Kafka is configured with default settings 
and the data (Section~\ref{ssec:data_source}) is streamed into two topics\footnote{Kafka topics: \url{https://developer.confluent.io/learn-kafka/apache-kafka/topics/}};
``\texttt{ndwFlow}'', and ``\texttt{ndwSpeed}'' containing the records about the number, and
the average speed of the cars respectively.

\section{Results}
\label{sec:results}
In this section, we discuss the results of our evaluation using different workloads.

\paragraph{Workload for throughput measurement}
For the throughput measurement workload, we ran the evaluation multiple times with
increasing input data throughput for each run to 
evaluate the sustainable throughput of the SUTs. 

In the first few runs of the evaluation, the RMLStreamer 
fared a bit worse than SPARQL-Generate in all 
three measurements. This is due to the overhead of having a distributed task manager
for executing, and managing the different 
tasks and subtasks of mapping heterogeneous data (Figure~\ref{fig:rml_architecture}).
However, when the 
throughput starts increasing beyond 10,000 records per second, RMLStreamer-SISO outperforms
SPARQL-Generate in terms of latency and memory usage.

\begin{figure}[!ht]
    \centering
    \includegraphics[width=\linewidth]{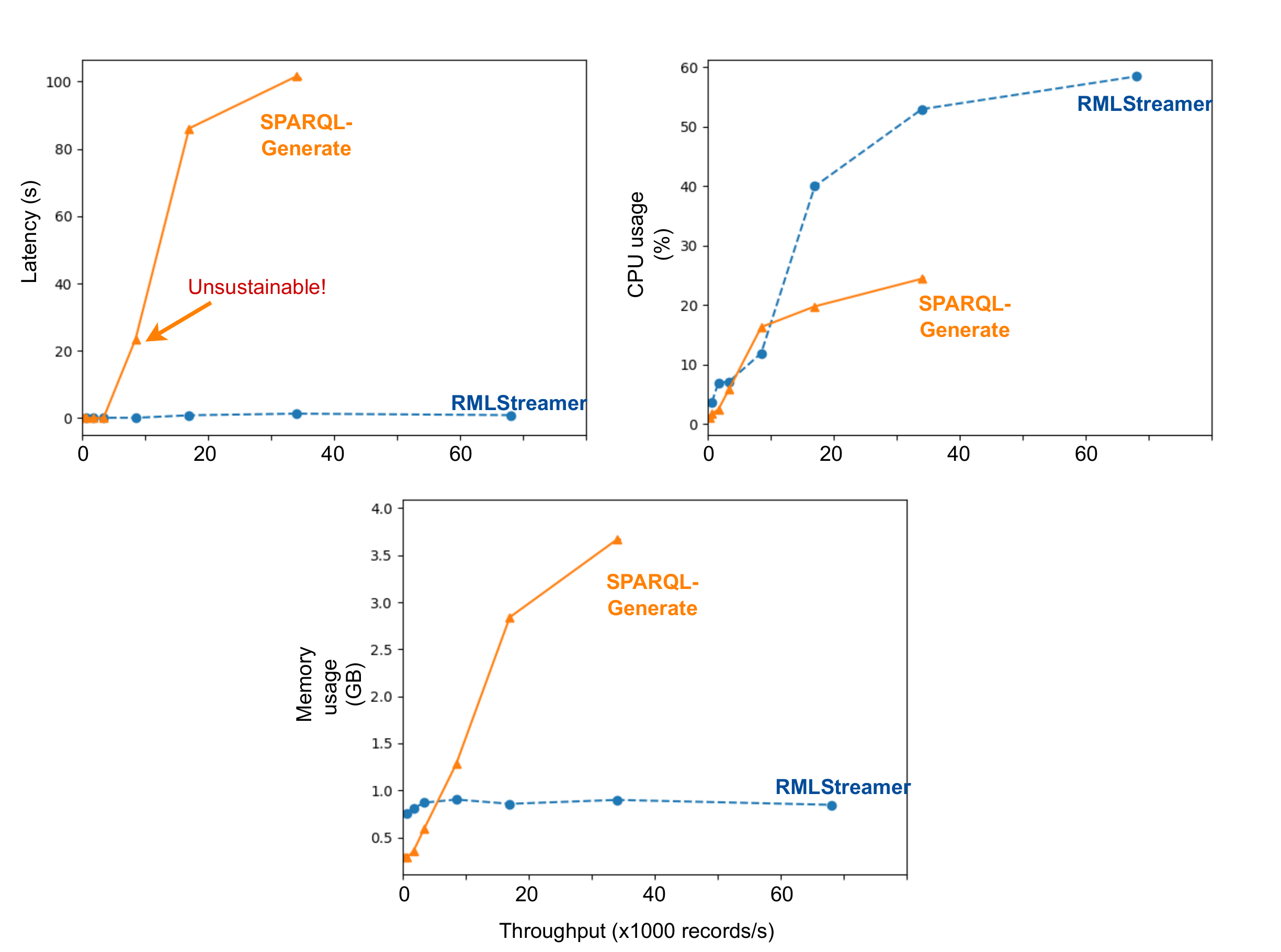}
    \caption{SUTs performance under different data stream velocity for 
    sustainable throughput measurement. 
    The last run for SPARQL-Generate was omitted because it took more than 1 hour instead of the expected 30 minutes to process the whole data stream.}
    \label{fig:throughput_result}
\end{figure}

Compared to RMLStreamer-SISO,
SPARQL-Generate became unsustainable when the throughput 
of the input data streams passes 10,000 
records per second with 20s latency (Figure~\ref{fig:throughput_result}). 
To the contrary, RMLStreamer-SISO has a consistent
low latency of 1 second for all runs of the workload 
having 100x magnitude lower latency than SPARQL-Generate 
in later runs.

Regarding CPU usage, RMLStreamer-SISO has on average 20\%
more CPU usage for the overhead of Apache Flink managing the 
distributed tasks.

In terms of memory usage, RMLStreamer-SISO uses
significantly lower memory than SPARQL-Generate 
at around 900MB compared to 3GB by SPARQL-Generate. 
Based on the previous observations,
we conclude that RMLStreamer-SISO outperforms SPARQL-Generate 
at higher throughput with lower latency and memory usage. Even
though, RMLStreamer-SISO's 
CPU usage is around 30\% higher than SPARQL-Generate in the last run, it effectively copes with
the increase in data stream velocity to maintain low latency processing.

\begin{figure}[!ht]
    \centering
    \includegraphics[width=\linewidth]{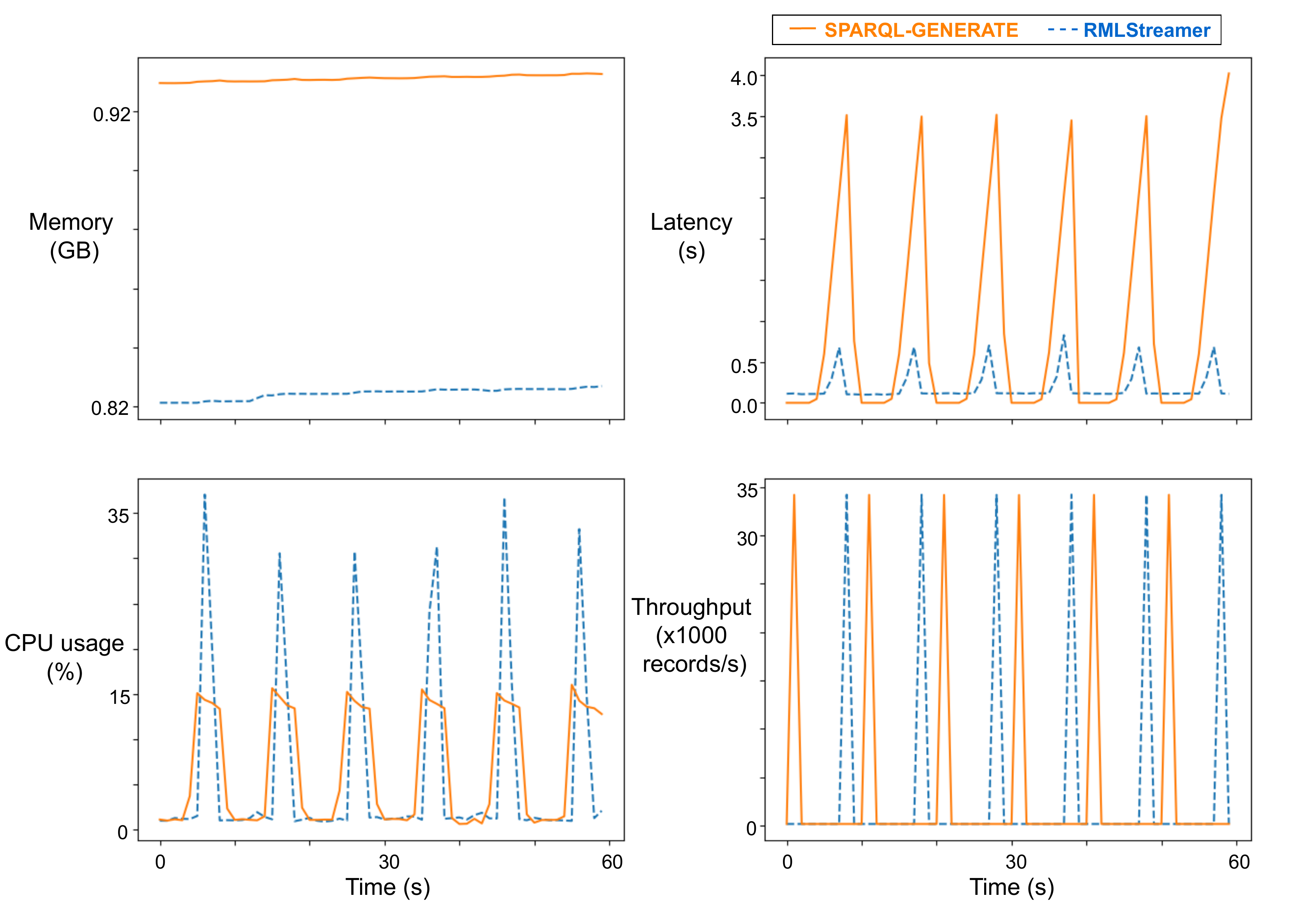}
    \caption{Performance of SUTs in the last one minute of the periodic burst workload evaluation. A part of the \textit{throughput} graph is blurred to give more clarity to the 
    relationship between the trends in \textit{latency} and \textit{throughput} of the engines.}
    \label{fig:periodic_result}
\end{figure}

\paragraph{Workload for periodic burst}
The periodic burst workload studies the adaptability of the
engine 
to the recurring sudden burst of data stream. We used the 
measurements from the last minute of the evaluation, when the
engines 
are \textit{stable} without warm-up overheads, to better
visualize 
their performance during the periodic burst of data
(Figure~\ref{fig:periodic_result}).
In Figure~\ref{fig:periodic_result}, we see a periodic increase,
and 
drop in the throughput metrics measurements, which is an
expected 
behaviour in the engines when consuming a data stream input with
periodic burst 
of data. Every 10 seconds we see a burst of around 35,000
messages. 
Both engines 
behave as expected for the throughput metrics measurement. 

The spikes for latency measurement of SPARQL-Generate (Figure~\ref{fig:periodic_result})
have a wider base than those of
RMLStreamer-SISO. This indicates that SPARQL-Generate takes a longer
time to recover from processing periodic workload 
than RMLStreamer-SISO by a few seconds.
RMLStreamer-SISO's peak latency is around 500ms whereas 
SPARQL-Generate has a peak latency of around $3.5s$.
Although RMLStreamer-SISO uses more CPU than
SPARQL-Generate to process data burst, 
it adapts to the sudden burst of data and recover more
quickly than SPARQL-Generate. 
The latency of RMLStreamer-SISO is also 7 times
lower than SPARQL-Generate due to the record-based 
processing capabilities.
We conclude that RMLStreamer-SISO is better adapted to workloads with periodic burst of data 
with faster recovery period, lower latency and memory usage while maintaining the same
throughput capabilities as SPARQL-Generate. 

\paragraph{Workload for scalability measurement} 

We evaluated the RMLStreamer-SISO's capability to join two data streams 
with a constant throughput of around 17000 messages per second. CPU and memory 
usage of both modes of RMLStreamer-SISO
are similar throughout the evaluation.  However,
despite the similar performance in terms of CPU and memory usage,
\textit{parallelized}
mode fared significantly better in terms of the latency
metric than 
\textit{unparallelized} mode. Unparallelized 
mode has a median latency 
of around 50000ms whereas 
\textit{parallelized} mode has a median latency of around
57ms. 
This is around 1000x lower 
in terms of the median latency. 
Moreover, the minimum latency of \textit{parallelized}
RMLStreamer-SISO at 8ms 
is 10,000x lower than the minimum latency of
\textit{unparallelized}
RMLStreamer-SISO at 13653ms. The latency is kept 
low with high
parallelization due to the 
effective distribution of the workload amongst the different
parallelized tasks 
by the underlying DSP engine (Apache Flink). 
We conclude that RMLStreamer scales extremely well 
with 
significantly better performance in terms of 
latency if
configured
to be executed in a distributed mode.

\section{Use Cases}
\label{sec:useCases}

RMLStreamer-SISO has seen uptake in multiple projects
--~covering different use cases in different architectures~--
to process streaming data and generate RDF streams.
Largest validation was in research and development (R\&D) projects between imec and Flemish companies
such as DyVerSIFy on streaming data analysis and visualisation~\cite{Steenwinckel2021FLAGSmethodologyadaptive,Paepe2021CompleteSoftwareStack}, together with Televic Rail on IoT data,
DAIQUIRI%
\footnote{\url{https://www.imec-int.com/en/what-we-offer/research-portfolio/daiquiri}}
together with VRT on sport sensor data,
and ESSENCE
and H2020 project MOS2S%
\footnote{\url{https://innovatie.vrt.be/project/essence}, \url{https://itea4.org/project/mos2s.html}} on media data.
Other projects include DiSSeCt%
\footnote{\url{https://smit.vub.ac.be/project/dissect}} on health data and transport data~\cite{Brouwer2020DistributedContinuousHome}.
The variety of use cases shows that the resource is suitable for solving the task at hand and also applicable to a multitude of use cases for society in general.
Applications --~within the knowledge graph construction problem domain~-- are varied, i.e.,
processing a large amount of low-frequency sensor data, a small amount of high-frequency sensor data, and large data sets combined with streaming data, processing Kafka streams, MQTT, Socket.io, and TCP streams.
Beyond Belgium, RMLStreamer has received attention by
the Institute of Data Science,
proposed as part of RDF graph generation tutorials such as those by STIInnsbruck in Austria, and
services such as Data2Services%
\footnote{\url{https://maastrichtu-ids.github.io/best-practices/blog/2021/03/18/build-a-kg/},\url{https://stiinnsbruck.github.io/lkgt/},\url{https://d2s.semanticscience.org/docs/d2s-rml/}} by the Institute of Data Science in Maastricht in the Netherlands.


\section{Conclusion and Future Work}
\label{sec:conclusions}

In this paper, we present RMLStreamer-SISO,
a highly scalable solution to seamlessly generate RDF streams
thanks to its dynamic window algorithm 
which adapts its window size to handle the dynamic characteristics of the data stream.
This way, RMLStreamer-SISO enables low latency and high
throughput mapping of heterogeneous data to RDF data.
We showed that our solution scales better than the state-of-the-art 
in terms of latency, memory, and throughput.
It is the only RDF stream generator 
which joins unbounded data streams and scale 
horizontally and vertically,
enabling RDF streams generation
from heterogeneous data streams
which was not possible so far.
Given it is open source and already widely used in different use cases
involving not only academia but also industry,
as shown in our use cases,
it is expected that the community that grew around it
will further grow and contribute at its maintenance,
while its extensive documentation and tutorials allow for easy reuse\footnote{Example of tutorial for use with docker technology, \url{https://github.com/RMLio/RMLStreamer/tree/development/docker}}.
The RML extensions will be further discussed
with the W3C community group on knowledge graph construction 
and eventually will be incorporated to the revised RML specification.

RMLStreamer-SISO increases the availability of RDF streams 
following the high availability of data streams.
Using a low-latency tool like RMLStreamer-SISO, 
legacy streaming systems could exploit the unique characteristics of real-life streaming data,
while enabling analysts to exploit the semantic reasoning using knowledge graphs in real-time.
This way, we enabled access to more data which should 
impact the further improvements of RSP engines
and other semantic web technologies on top of RDF streams
which were not possible so far.





\paragraph*{Resource Availability Statement:}
Source code for RMLStreamer-SISO is available at 
\url{https://github.com/RMLio/RMLStreamer}.
The source code for the benchmark is available at \url{https://github.com/s-minoo/rmlstreamer-benchmark-rust}.
The dataset used for the benchmark is available at 
\url{https://github.com/Klarrio/open-stream-processing-benchmark/tree/master/data-stream-generator}.

%
%
\bibliographystyle{splncs04}
\bibliography{main}

\begin{thebibliography}{10}
\providecommand{\url}[1]{\texttt{#1}}
\providecommand{\urlprefix}{URL }
\providecommand{\doi}[1]{https://doi.org/#1}

\bibitem{c-sparql}
Barbieri, D.F., Braga, D., Ceri, S., Della~Valle, E., Grossniklaus, M.:
  C-sparql: Sparql for continuous querying. In: Proceedings of the 18th
  International Conference on World Wide Web. p. 1061–1062. WWW '09,
  Association for Computing Machinery, New York, NY, USA (2009).
  \doi{10.1145/1526709.1526856}

\bibitem{chimera_ontop}
Belcao, M., Falzone, E., Bionda, E., Valle, E.D.: Chimera: A bridge between big
  data analytics and semantic technologies. In: Hotho, A., Blomqvist, E.,
  Dietze, S., Fokoue, A., Ding, Y., Barnaghi, P., Haller, A., Dragoni, M.,
  Alani, H. (eds.) The Semantic Web -- ISWC 2021. pp. 463--479. Springer
  International Publishing, Cham (2021)

\bibitem{survey_dsp}
Botan, I., Derakhshan, R., Dindar, N., Haas, L., Miller, R.J., Tatbul, N.:
  Secret: A model for analysis of the execution semantics of stream processing
  systems. Proc. VLDB Endow.  \textbf{3}(1–2),  232–243 (sep 2010).
  \doi{10.14778/1920841.1920874}

\bibitem{Brouwer2020DistributedContinuousHome}
Brouwer, M.D., Bonte, P., Arndt, D., Sande, M.V., Heyvaert, P., Dimou, A.,
  Verborgh, R., Turck, F.D., Ongenae, F.: Distributed continuous home care
  provisioning through personalized monitoring {\&} treatment planning. In:
  Companion Proceedings of the Web Conference 2020. {ACM} (Apr 2020).
  \doi{10.1145/3366424.3383528}

\bibitem{sparql_stream}
Calbimonte, J.P., Corcho, O., Gray, A.J.G.: Enabling ontology-based access to
  streaming data sources. In: Patel-Schneider, P.F., Pan, Y., Hitzler, P.,
  Mika, P., Zhang, L., Pan, J.Z., Horrocks, I., Glimm, B. (eds.) The Semantic
  Web -- ISWC 2010. pp. 96--111. Springer Berlin Heidelberg, Berlin, Heidelberg
  (2010)

\bibitem{flink}
Carbone, P., Katsifodimos, A., Ewen, S., Markl, V., Haridi, S., Tzoumas, K.:
  Apache flink{\texttrademark}: Stream and batch processing in a single engine.
  IEEE Data Eng. Bull.  \textbf{38},  28--38 (2015)

\bibitem{chiu1989analysis}
Chiu, D.M., Jain, R.: Analysis of the increase and decrease algorithms for
  congestion avoidance in computer networks. Computer Networks and ISDN systems
   \textbf{17}(1),  1--14 (1989)

\bibitem{fno}
De~Meester, B., Dimou, A., Verborgh, R., Mannens, E.: An ontology to
  semantically declare and describe functions. In: Sack, H., Rizzo, G.,
  Steinmetz, N., Mladeni{\'{c}}, D., Auer, S., Lange, C. (eds.) The Semantic
  Web. pp. 46--49. Springer International Publishing, Cham (2016)

\bibitem{dsp_edge}
{Dias de Assunção}, M., {da Silva Veith}, A., Buyya, R.: Distributed data
  stream processing and edge computing: A survey on resource elasticity and
  future directions. Journal of Network and Computer Applications
  \textbf{103},  1--17 (2018). \doi{10.1016/j.jnca.2017.12.001},
  \url{https://www.sciencedirect.com/science/article/pii/S1084804517303971}

\bibitem{rml}
Dimou, A., Vander~Sande, M., Colpaert, P., Verborgh, R., Mannens, E., Van~de
  Walle, R.: Rml: A generic language for integrated rdf mappings of
  heterogeneous data. vol.~1184 (04 2014)

\bibitem{dsp_eval}
van Dongen, G., Van~den Poel, D.: Evaluation of stream processing frameworks.
  IEEE Transactions on Parallel and Distributed Systems  \textbf{31}(8),
  1845--1858 (2020). \doi{10.1109/TPDS.2020.2978480}

\bibitem{generic}
Gedik, B.: Generic windowing support for extensible stream processing systems.
  Softw. Pract. Exper.  \textbf{44}(9),  1105–1128 (sep 2014).
  \doi{10.1002/spe.2194}

\bibitem{rmlstreamer-big-data}
Haesendonck, G., Maroy, W., Heyvaert, P., Verborgh, R., Dimou, A.: Parallel rdf
  generation from heterogeneous big data. In: Proceedings of the International
  Workshop on Semantic Big Data. SBD '19, Association for Computing Machinery,
  New York, NY, USA (2019). \doi{10.1145/3323878.3325802}

\bibitem{sdm-rdfizer}
Iglesias, E., Jozashoori, S., Chaves-Fraga, D., Collarana, D., Vidal, M.E.:
  Sdm-rdfizer. Proceedings of the 29th ACM International Conference on
  Information \& Knowledge Management  (Oct 2020).
  \doi{10.1145/3340531.3412881}

\bibitem{karimov_2018}
Karimov, J., Rabl, T., Katsifodimos, A., Samarev, R., Heiskanen, H., Markl, V.:
  Benchmarking distributed stream data processing systems. 2018 IEEE 34th
  International Conference on Data Engineering (ICDE)  (Apr 2018).
  \doi{10.1109/icde.2018.00169}

\bibitem{c-qels}
Le~Phuoc, D., Dao-Tran, M., Le~Tuan, A., Duc, M.N., Hauswirth, M.: Rdf stream
  processing with cqels framework for real-time analysis. In: Proceedings of
  the 9th ACM International Conference on Distributed Event-Based Systems. p.
  285–292. DEBS '15, Association for Computing Machinery, New York, NY, USA
  (2015). \doi{10.1145/2675743.2772586}

\bibitem{sparql-gen}
Lefran{\c{c}}ois, M., Zimmermann, A., Bakerally, N.: A sparql extension for
  generating rdf from heterogeneous formats. In: Blomqvist, E., Maynard, D.,
  Gangemi, A., Hoekstra, R., Hitzler, P., Hartig, O. (eds.) The Semantic Web.
  pp. 35--50. Springer International Publishing, Cham (2017)

\bibitem{triplewave}
Mauri, A., Calbimonte, J.P., Dell'Aglio, D., Balduini, M., Brambilla, M.,
  Della~Valle, E., Aberer, K.: Triplewave: Spreading rdf streams on the web.
  In: Groth, P., Simperl, E., Gray, A., Sabou, M., Kr{\"o}tzsch, M., Lecue, F.,
  Fl{\"o}ck, F., Gil, Y. (eds.) The Semantic Web -- ISWC 2016. pp. 140--149.
  Springer International Publishing, Cham (2016)

\bibitem{storm}
N.A: Apache storm, \url{https://storm.apache.org/}

\bibitem{Paepe2021CompleteSoftwareStack}
Paepe, D.D., Hautte, S.V., Steenwinckel, B., Moens, P., Vaneessen, J.,
  Vandekerckhove, S., Volckaert, B., Ongenae, F., Hoecke, S.V.: A complete
  software stack for {IoT} time-series analysis that combines semantics and
  machine learning{\textemdash}lessons learned from the dyversify project.
  Applied Sciences  \textbf{11}(24),  11932 (dec 2021).
  \doi{10.3390/app112411932}

\bibitem{rdf-gen}
Santipantakis, G.M., Kotis, K.I., Vouros, G.A., Doulkeridis, C.: Rdf-gen:
  Generating rdf from streaming and archival data. In: Proceedings of the 8th
  International Conference on Web Intelligence, Mining and Semantics. WIMS '18,
  Association for Computing Machinery, New York, NY, USA (2018).
  \doi{10.1145/3227609.3227658}

\bibitem{chimera}
Scrocca, M., Comerio, M., Carenini, A., Celino, I.: Turning transport data to
  comply with eu standards while enabling a multimodal transport knowledge
  graph. The Semantic Web – ISWC 2020 p. 411–429 (2020).
  \doi{10.1007/978-3-030-62466-8_26}

\bibitem{Simsek2019RocketRMLA}
Simsek, U., K{\"a}rle, E., Fensel, D.A.: Rocketrml - a nodejs implementation of
  a use case specific rml mapper. ArXiv  \textbf{abs/1903.04969} (2019).
  \doi{10.48550/ARXIV.1903.04969}

\bibitem{Steenwinckel2021FLAGSmethodologyadaptive}
Steenwinckel, B., Paepe, D.D., Hautte, S.V., Heyvaert, P., Bentefrit, M.,
  Moens, P., Dimou, A., Bossche, B.V.D., Turck, F.D., Hoecke, S.V., Ongenae,
  F.: {FLAGS}: A methodology for adaptive anomaly detection and root cause
  analysis on sensor data streams by fusing expert knowledge with machine
  learning. Future Generation Computer Systems  \textbf{116},  30--48 (Mar
  2021). \doi{10.1016/j.future.2020.10.015}

\bibitem{rsplab}
Tommasini, R., Della~Valle, E., Mauri, A., Brambilla, M.: Rsplab: Rdf stream
  processing benchmarking made easy. In: d'Amato, C., Fernandez, M., Tamma, V.,
  Lecue, F., Cudr{\'e}-Mauroux, P., Sequeda, J., Lange, C., Heflin, J. (eds.)
  The Semantic Web -- ISWC 2017. pp. 202--209. Springer International
  Publishing, Cham (2017). \doi{10.1007/978-3-319-68204-4_21}

\bibitem{spark}
Zaharia, M., Xin, R.S., Wendell, P., Das, T., Armbrust, M., Dave, A., Meng, X.,
  Rosen, J., Venkataraman, S., Franklin, M.J., Ghodsi, A., Gonzalez, J.,
  Shenker, S., Stoica, I.: Apache spark: A unified engine for big data
  processing. Commun. ACM  \textbf{59}(11),  56–65 (oct 2016).
  \doi{10.1145/2934664}

\bibitem{adaptive}
Zhang, Q., Song, Y., Routray, R.R., Shi, W.: Adaptive block and batch sizing
  for batched stream processing system. In: 2016 IEEE International Conference
  on Autonomic Computing (ICAC). pp. 35--44 (2016). \doi{10.1109/ICAC.2016.27}

\end{thebibliography}
\end{document}